\documentclass[aps,11pt]{revtex4}
\usepackage{latexsym}
\begin{document}
\newcommand{\beq}{\begin{equation}}
\newcommand{\eeq}{\end{equation}}
\renewcommand{\baselinestretch}{1.5}
\newcommand{\tif}{\tilde{F}}
\newcommand{\cl}{{\cal L}}
%
%
\newcommand{\Prd}{Phys.  Rev D}
\newcommand{\Prl}{Phys.  Rev.  Lett.}
\newcommand{\Plb}{Phys.  Lett.  B}
\newcommand{\Cqg}{Class.  Quantum Grav.}
\newcommand{\Np}{Nuc.  Phys.}
\newcommand{\Grg}{Gen. Rel. \& Grav.}
\newcommand{\Fp}{Fortschr. Phys.}

\renewcommand{\baselinestretch}{1.2}


\title{Propagation of perturbations in non-linear spin-2 theories}

\author {Santiago E.  Perez Bergliaffa}

\affiliation{Centro Brasileiro de Pesquisas Fisicas, Rua Dr. Xavier Sigaud, 150, CEP
22290-180, Rio de Janeiro, Brazil}

\date{\today}
\begin{abstract}
In this communication I analyze the problem of complete
exceptionality of wave propagation in a class of spin 2 field
theories. I show that, under the imposition of the good weak-field
limit, only two Lagrangians are completely exceptional. These are
the linear Fierz Lagrangian, and a
Born-Infeld-like Lagrangian. As a byproduct, I reobtain the result
that in a nonlinear theory, spin 2 particles follow an effective
metric that depends on the nonlinearities of the Lagrangian.
\end{abstract}
\maketitle
\vskip2pc

It is a pleasure for me to contribute to this volume to
celebrate Mario Novello's 60th birthday. Amongst the several
areas of research in General Relativity (GR), Cosmology, and Field
Theory that Mario has contributed to, the physics of non-linear
spin 2 field theories has remained, I believe, quite unexplored.
This is an area that deserves some attention: although GR has
passed many observational tests with flying colours, all of them
are carried out in the weak-field limit. In other words, there are
no tests of the behaviour of gravity in the strong field regime.
And it is precisely in this regime where the non-linearities are
expected to be important. Lacking any experimental evidence,
it would be important to restrict the theory space by
some other means. If we start from a {\em generic} theory of
gravitation and impose {\em plausible} assumptions, can we find a
{\em class} of theories that agree with GR in the situations where
the latter has been tested but with a different behavior in
strong-field situations? To put it another way, how "general" is
GR?

A possible way to start to build an answer to this question would
be take the most general gravitational Lagrangian $\cl$, and in a
first step, restrict its form as much as possible by the
imposition of sound physical principles. A second step would be to
confront the predictions obtained from this Lagrangian with the
available observations. We could use the standard geometrical
approach to gravitation (``Einstein representation''), but a good
amount of work in the ``squeezing the theory space'' has already
been done in Electromagnetism \cite{deser1,deser2}. We can benefit
from this work if we use the Fierz-Pauli (FP) formulation  of spin
2 theories \cite{fp,nn}. In this contribution I will examine the
propagation of waves in non-linear spin 2 field theories using the
FP representation. In particular, I will show that out of a
certain class of spin 2 nonlinear theories, only two Lagrangians
satisfy the physical requirement of complete exceptionality ({\em
i.e.} absence of shocks in wave propagation). But let me begin
first by a brief summary of the FP representation.

\section{FP representation of spin 2 fields}

In 1939, Fierz and Pauli showed that a spin 2 field can be
represented by a third rank tensor $F_{\alpha\mu\nu}$ with the
properties \cite{fp}:
$$
F_{\alpha\mu\nu} + F_{\mu\alpha\nu} = 0,
$$
$$
F_{\alpha\mu\nu} + F_{\mu\nu\alpha} + F_{\nu\alpha\mu} = 0.
$$
The dual of the object $F$ is defined by:
\[
\tif^{\alpha\mu}{}_{\lambda} \equiv \frac{1}{2} \,
\eta^{\alpha\mu}{}_{\nu\sigma}\,F^{\nu\sigma}{}_{\lambda}.
\]
We shall impose the constraint
$$
\tif^{\alpha (\mu\nu)}{}_{,\alpha} = 0,
$$
which guarantees that $F_{\mu\nu\alpha}$ represents a spin-two
field \cite{nn} (the comma means derivative in the flat background).
It also allows the introduction of a
``potential'' $\varphi_{\mu\nu}$:
$$
F_{\alpha\mu\nu} = \frac 1 2 \left( \varphi_{\nu [\alpha ,\mu ]} +
\varphi_{[,\alpha} \eta_{\mu]\nu} + \eta_{\nu[\alpha}
\varphi^{\lambda}_{\;\mu],\lambda} \right) .
$$
Taking the trace, we get
$$
F_{\alpha} = \varphi_{,\alpha} -
\varphi_{\alpha}{}^{\lambda}{}_{,\lambda}.
$$
$F_{\mu\nu\alpha}$ satisfies
$$
F^{\alpha }{}_{(\mu \nu ),\alpha }\equiv -\,G^{L}{}_{\mu \nu },
$$
where
$$
G^{L}{}_{\mu \nu }\equiv \Box \,\varphi _{\mu \nu }-\varphi
^{\epsilon }{}_{(\mu ,\nu )\,,\epsilon }+\varphi _{,\mu \nu }-\eta
_{\mu \nu }\,\left( \Box \varphi -\varphi ^{\alpha \beta
}{}_{,\alpha \beta }\right)
$$
is the linearized Einstein tensor.

In the weak-field limit, we expect $\cl$ to agree with linearized
GR, so in this regime the EOM for the Fierz-Pauli tensor must be
$$
F^{\alpha }{}_{(\mu \nu ),\alpha } =0,
$$
which is obtainable from the Lagrangian
$$
\cl = x-y,$$ where
$$
x \equiv F_{\alpha \mu \nu }\hspace{0.5mm}F^{\alpha \mu \nu },
\;\;\;\;\;\;\;\;\;\;\;\;
y \equiv F_{\mu }\hspace{0.5mm}F^{\mu }.
$$
We can define yet another invariant:
$$
w \equiv F_{\alpha \beta \lambda }\tif^{\alpha \beta \lambda
}=\frac{1}{2}\,F_{\alpha \beta \lambda }\hspace{0.5mm}F^{\mu \nu
\lambda }\,\eta ^{\alpha \beta }{}_{\mu \nu },
$$
which is not a topological invariant if the Lagrangian is
nonlinear. In what follows, we shall consider Lagrangians of the
form
$$
\cl = \cl (z, w^2)
$$
(with $z=x-y$), as suggested by the weak-field limit and to ensure
parity invariance. The EOM that follow from this Lagrangian are
\beq
\left[ \cl_ zF^{\alpha(\delta\chi)} +2w \tif^{\alpha(\delta\chi)}
\right]_{;\alpha}
-2\;\gamma^{\delta\chi}(w\cl_{w^2}\tif^\alpha)_{;\alpha}+
+2(w\cl_{w^2} \tif^{(\delta}\;)_{;\lambda}
\;\gamma^{\chi)\lambda} =  0,
\eeq
where $\tif^\alpha \equiv \gamma_{\mu\nu}\tif^{\alpha\mu\nu}$, and
$\gamma^{\delta\chi}$ is the metric of flat spacetime. We shall
introduce in the next section one of the physical principles that
will restrict the form of ${\cl}$.

\section{Propagation of the discontinuities}

{\em Shocks} are unwanted features of certain nonlinear systems. A
shock is an infinite discontinuity of the field, where by
``infinite discontinuity'' we actually mean a large change in the
field value in a very short distance. Shocks form whenever wave
fronts ``pile up'', as in the case of {\em caustics} in optics.
However, these are due to external causes (like dielectric media).
Here instead if there are shocks, they originate from the dynamics
of the theory {\em in vacuo}. Notice also that the EOM describing
the field are not valid in the region where a shock is present. We
shall start then by limiting the possible Lagrangians by imposing
"good propagation" ({\em i.e.} absence of shocks) for the spin-2
field, following the seminal work of Boillat \cite{boillat} on
propagation of waves in nonlinear electrodynamics. Let us begin
with some conventions and notation. We shall assume that across
the hypersurface $S$ (the wave surface), given by the equation
$$
\phi(x^\alpha) = 0,\;\; \alpha = 0,1,...,n,
$$
the highest derivatives of the dependent field variables that
appear in the field equations are discontinuous. In order to state
more precisely this hipothesis, it is convenient to adopt new
coordinates:
$$
\{x^\alpha\} \rightarrow \{\phi(x^\alpha ),\; \xi^i(x^\alpha )\},
\;\;i=1,2,...,n.
$$
Our assumption is then that the jump through $S$ of the order-$q$
derivative (which is the highest derivative in the E.O.M) of the
field component $u$, given by \beq \left[
\frac{\partial^qu}{\partial\phi^q}\right] =
\left(\frac{\partial^qu}{\partial\phi^q} \right)_{\phi =0^+} -
\left(\frac{\partial^qu}{\partial\phi^q} \right)_{\phi =0^-}
\equiv \delta^q u , \label{disc} \eeq is finite, while $\delta^r u
=0,\;\;0\leq r < q$.

In order to allow for the discontinuities given in
Eqn.(\ref{disc}), $\phi (x^\alpha)$ must be a solution of a
characteristic equation of the form \beq H \equiv G^{\alpha\beta
...\nu}\phi_\alpha \phi_\beta ...\phi_\nu = 0 \label{hamil} \eeq
where the completely symmetric tensor $G$ may depend on the field
and all its continuous derivatives in the case of nonlinear
theories, and $\phi_\alpha \equiv \partial\phi/\partial x^\alpha$
\cite{boillat2}.

The general theory of wave propagation shows that given an E.O.M.,
typically there exist several modes of propagation, and for each
of them the wave surface moves with a different normal velocity.
These velocities are given by $v^{(i)}_n = \lambda^{(i)} \vec n$,
where the $\lambda^{(i)}$ are the eigenvalues of a matrix
associated to the E.O.M., and $\vec n$ is the unit vector normal
to $S$
 \cite{boillat2}. A {\em shock} ({\em i.e.}
a jump in the field itself, which implies the divergence of its
normal derivative) forms whenever wave fronts of a given
propagation mode ``pile up''. A sufficient condition for the
avoidance of shocks for a given mode is then that the velocity of
the wave front be independent of the coordinate $\phi$
\cite{boillat2,deser2}. That is to say, \beq \left[\partial_\phi
\lambda^{(i)}\right] =0 \eeq When this condition is satisfied, it
is said that the corresponding wave is exceptional. If this
condition is satisfied by all the eigenvalues $\lambda^{i}$, then
the system is said to be {\em completely exceptional} (CE). The
condition for a system to be CE can be also written as
\cite{boillat,deser2} \beq \delta H \equiv \phi_\alpha \phi_\beta
...\phi_\nu \;\delta G^{\alpha\beta ...\nu}= 0. \label{ce} \eeq It
can be shown that electromagnetic waves in Maxwell theory are CE,
as well as gravitational waves in Einstein's theory \cite{deser2}.
Shocks are present, however, in the nonlinear electromagnetic
theory described by the Euler-Heisenberg Lagrangian \cite{euheis}.

\section{ Application of the method to spin-2 nonlinear
theories}

To use the sufficient condition given in the previous section, we
need to determine the function H. Let's go back to the particular
case of spin-2 fields with a Lagrangian
$$
\cl = \cl (z, w^2),
$$
with $z=x-y$. The EOM are
\beq
\left[ \cl_ zF^{\alpha(\delta\chi)} +2w \tif^{\alpha(\delta\chi)}
\right]_{;\alpha}
-2\;\gamma^{\delta\chi}(w\cl_{w^2}\tif^\alpha)_{;\alpha}+
+2(w\cl_{w^2} \tif^{(\delta}\;)_{;\lambda}
\;\gamma^{\chi)\lambda} = 0.
\eeq
We shall assume that
$$
\delta^{(2)}\varphi_{\alpha\beta} \equiv \pi_{\alpha\beta} \neq 0,
$$
From the expression for the field F,
$$
F_{\alpha\mu\nu} = \frac 1 2 \left( \varphi_{\nu [\alpha ,\mu ]} +
\varphi_{[,\alpha} \eta_{\mu]\nu} + \eta_{\nu[\alpha}
\varphi^{\lambda}_{\;\mu],\lambda} \right),
$$
we see that its first derivative is discontinuous.

The following definitions will be useful in what follows:
$$
V_{\mu\nu} = \phi^\lambda F_{\lambda\mu\nu},\;\;\;\;\;\;\;\;\;\;
M_{\delta\chi} = \phi_{\delta}F_\chi,
$$
$$
Z_{\delta\chi} = \phi_\alpha
\tif^\alpha_{\;\delta\chi},\;\;\;\;\;\;\;\;\;\; N_{\delta\chi} =
\phi_\chi \tif_\delta,
$$
$$
U_\alpha = \phi_\nu \pi^\nu_\alpha.
$$
With the help of the rule
$$
\delta^{(0)}u_{;\alpha} = \phi_\alpha\; \delta^{(1)}u,
$$
we can take the discontinuity of the E.O.M. Omitting the index
$(1)$ from now on, we get the equations
\beq
V^{(\delta\chi)}\; \delta\cl_z + 2
\;(w\;\delta\cl_{w^2}+\cl_{w^2}\;\delta w) \;\left[
W^{(\delta\chi)} -
 (\phi . \tif )\;\gamma^{(\delta\chi)} +
N^{(\delta\chi)}\right] + \cl_z \;\phi_\alpha \;\delta
F^{\alpha(\delta\chi)}
 = 0.
\label{deom}
\eeq
The variations in this equation are given by
$$
\delta\cl_z = \delta\cl_{zz} \delta z + 2w\delta\cl_{zw^2}\delta
w,
$$
(and a similar equation for $\delta \cl_{w^2}$),
$$
\delta z = -2 V.\pi,
$$
$$
\delta w = 2(W.\pi - \pi (\phi . \tif) + U . \tif),
$$
$$
\phi_\alpha \delta F^{\alpha(\delta\chi)} =
\phi^{(\delta}U^{\chi)} - \phi^2 \;\pi^{\delta\chi} +
\;\gamma^{\delta\chi}\; (\pi \phi^2 - \phi.U) - \pi \phi^\delta
\phi^\chi.
$$
From Eqn.(\ref{deom}) we conclude that
$$
\pi ^{\delta\chi} = a V^{(\delta\chi)} + b W^{(\delta\chi)} + c
\gamma^{\delta\chi} + d N^{(\delta\chi)} + e \phi^\delta \phi^\chi
+ f \phi^{(\delta}U^{\chi)}.
$$
Comparing this expression with Eqn.(\ref{deom}), we can get a system of algebraic
equations for the coefficients, and obtainfrom it the function $H$.
The calculations are rather clumsy, so we shall restrict here to the case
$$
\cl = \cl (z),
$$
for which $W^{(\delta\chi)}= N^{(\delta\chi)}=0$, and so
$$
\pi ^{\delta\chi} = a\; V^{(\delta\chi)} +  c
\;\gamma^{\delta\chi} + e \;\phi^\delta \phi^\chi + f\;
\phi^{(\delta}U^{\chi)}.
$$
Projecting Eqn.(\ref{deom}) with the tensors appearing in
$\pi^{\delta\chi}$, we get after some algebra the characteristic
equation
$$
H = \phi_\alpha\phi_\beta\left[ \gamma^{\alpha\beta} - 2
\;\frac{\cl_{zz}}{\cl_z} \left(F^\alpha F^\beta -2
F^{\alpha\rho\sigma}F^\beta_{\;\;\rho\sigma}\right)\right]=0.
$$
This is the ``effective metric'' obtained before in Novello, De
Lorenci, and de Freitas \cite{ndl1}. It shows that gravitons do not
move on the light cone, a prediction that will be tested in the
near future by  gravitational-wave observatories.

Notice that if $\cl _{zz} =0$, the velocity of the gravitational waves will
coincide with that of electromagnetic waves. In analogy to
what happens in nonlinear electromagnetism \cite{novem},
in the more general case $\cl =\cl(z,w^2)$ we expect that
$$
H =\phi_\alpha\phi_\beta\left[ \gamma^{\alpha\beta} +
A^{\alpha\beta}\right]
$$
where
$$
A^{\alpha\beta} = A^{\alpha\beta}(\cl_z, \cl_{zz},
\cl{zw^2},\cl_{w^2},\cl_{w^2w^2},F).
$$
Imposing that $c_{gw} =c_{EM}$ would give another
differential constraint to be satisfied by $\cl$.

\section{Completely Exceptional Lagrangians}

As we mentioned before, complete exceptionality is the property that
guarantees that initial wavefronts evolve so as to prevent the
emergence of shocks.

The condition for CE was that $\delta H =0$. Taking the variation
of
$$
H = \phi_\alpha\phi_\beta\left[ \gamma^{\alpha\beta} - 2\;
\frac{\cl_{zz}}{\cl_z} \left(F^\alpha F^\beta -2
F^{\alpha\rho\sigma}F^\beta_{\;\;\rho\sigma}\right)\right]=0,
$$
we get a differential equation for the Lagrangian $\cl (z)$ that
governs the shock-free evolution: \beq 3({\cal L}^{''})^2 - {\cal
L}^{'} {\cal L}^{'''} = 0. \eeq There are only two solutions of
this equation. These are the Fierz-Pauli Lagrangian, \beq {\cal
L}(z) = a z + b, \eeq and a Born-Infeld-like Lagrangian, \beq
{\cal L}(z) = \;\pm a \left( \sqrt{1 - \frac z b } \mp c \right),
\label{bil} \eeq where $a,b$ and $c$ are integration constants.

Only these two Lagrangians display the property of complete
exceptionality, in close parallel to the case of nonlinear
electromagnetism, with a Lagrangian that depends only on the
invariant $F_{\mu\nu}F^{\mu\nu}$ \cite{deser1}.

Let me mention that the theory defined by Eqn.(\ref{bil}) (known as NDL theory)
was studied in a series of
papers by M. Novello
and collaborators \cite{ndl1}.

\section{Discussion}

Let me summarize what was achieved here. First,
we reobtained, for the case $\cl = \cl (z)$, the result that
the velocity of gravitational waves will be in general different
from $c_{EM}$. This prediction will be put to test by the
observation of gravitational waves. Note however that it depends
on the background fields, through the specific choice of $\cl$.

Second, it was shown that in the simple case of $\cl = \cl (z)$, the
method we suggested in the Introduction works:
we started with a general Lagrangian, and the
imposition of a {\em single} ``sensible'' physical requirement
reduced drastically the theory space.

Third, the calculations for the more general case $\cl = \cl
(z,w^2)$ (under way) are more involved, but due to the use of the
Fierz-Pauli representation, they are formally similar to the
analogous case in Electromagnetism, and we can expect similar results.
For this general case, most probably other requirements will
be needed in addition to that of CE, like $c_{gw} = c_{EM}$,
positivity of the energy (or energy conditions?), ``duality''...

Finally, the allowed polarization states in these theories deserve
a detailed analysis. If
the number of states is to agree with that of GR (this assertion will be
tested in the near future by gravitational-waves astronomy), then we
would have another constraint to be satisfied by $\cl$.

\end{document}